# The next frontier in exoplanet science:
# Imaging our neighbouring planetary systems

*Ignas Snellen, Sebastiaan Haffert, Matthew Kenworthy, Tomas Stolker*
*Leiden Observatory, Leiden University, Postbus 9513, 2300 RA Leiden, The Netherlands*

*Transmission and eclipse spectroscopy have been invaluable tools for the characterisation of extrasolar planet atmospheres. While they will continue to provide many new insights and discoveries in the decade(s) to come, these methods are running up against sources of stellar noise from stellar surface inhomogeneities and variability. In this white paper we discuss how the next steps in the characterisation of small, temperate rocky planets requires high-contrast imaging, making the planetary systems around our closest neighbouring stars the new frontier in exoplanet science. The Extremely Large Telescopes (ELTs) will be at the forefront of this quest. The Planetary Camera & Spectrograph (PCS) on ESO's ELT and GmagAO-X on the GMT are planned to become operational in the 2035-2040 time-frame, allowing the characterisation of up to dozen(s) of rocky planets around nearby red dwarf stars. We discuss what role there will be still to play for ground-based exoplanet characterisation in the era of the space-borne Habitable Worlds Observatory and LIFE missions.*

**Exoplanet characterization: 1995 – 2025**

If there is one thing that we have learned from three decades of exoplanetary studies, it is that exoplanets are both very common and diverse. They appear in orbits with periods of under a day out to distances of hundreds of astronomical units from their parent star, with masses and sizes implying a wide range of compositions. Clearly, our Solar System is not a universal model for planetary systems[1]. Understanding what drives this diversity will also inform us about the worlds in our own Solar System: what makes our Earth unique? Are there any other planets with habitable conditions, and if so, do they harbour life? These fundamental questions are driving exoplanet research, and atmospheric measurements are a crucial element of this endeavour.

Exoplanet atmospheric characterisation started in earnest with HST in 2002 with the detection of sodium in the atmosphere of a transiting hot Jupiter[2]. Transmission spectroscopy, secondary eclipse measurements, and phase-curve observations have dominated atmospheric studies ever since[3]. In the pre-JWST era these were largely constrained to hot Jupiters through measuring their day-to-nightside heat distributions, vertical temperature structures, abundances of spectroscopically active molecules, and signatures of atmospheric escape. High-resolution spectroscopy discerns global wind patterns and spin-rotation signatures, and a wide range of atomic and ionic species in the most-irradiated atmospheres[4].

The unprecedented stability and sensitivity of JWST is revolutionizing atmospheric characterization[5], moving towards smaller and/or cooler planets; warm mini-Neptunes that appear to be within a new chemical regime, cloud-enshrouded super-Earths, and lava-worlds with volatile atmospheres. JWST is now vigorously pursuing dayside-temperature measurements of rocky exoplanets transiting M-dwarfs. The most informative observations are from the inner planets of the TRAPPIST-1 system, which are consistent with bare rock[6] surfaces (but do not exclude thin atmospheres).

The question is whether these techniques can make much further progress. By their nature, these time-differential methods have to cope with the full disk of stellar noise from primary stars that are many orders of magnitude brighter than the targeted planets. In addition, stars are not perfect calibrators; they exhibit intrinsic flux variations and surface inhomogeneity (i.e. spots, plages, and polar-to-equatorial temperature gradients) which produce an astrophysical noise floor that is extremely challenging to overcome[7]. The only way forward for detailed characterisation of temperate rocky planets and gas giants is by angularly separating the planets from their host stars on the sky through direct imaging.

**The next frontier: high-contrast imaging of nearby planetary systems**

High-contrast imaging started with the advent of adaptive optics systems on ground-based telescopes and with the HST in space. They discovered the first brown dwarfs at arcsecond wide separations[8]. However, since the diffraction limit is inversely proportional to telescope diameter, the current 8-10m class ground-based telescopes generally outperform HST at small angular separations. The first super-Jupiters, gas giant planets that straddle the planet/brown-dwarf divide in terms of their mass, were found in the mid-2000s, showing spectra rich in molecular absorption features. These imaged planets are all young (<100 Myr) and are still warm from their formation. This is what makes them detectable. More recently the discovery of accreting proto-planets, still embedded in their proto-planetary disks, gives direct observational evidence of the planet formation processes[9].

Improvements in direct imaging (high-contrast imaging; HCI), are largely driven by advancements in adaptive optics and coronagraphic techniques[10]. State-of-the-art instrumentation include SPHERE on ESO's VLT and GPI on Gemini. Great strides forward are being achieved with the GRAVITY instrument on the VLTI, which provides unprecedented angular resolution. However, detection of rocky planets with optical interferometry will be challenging as a finely sampled uv-plane is needed to reach the required dynamical range.

In its theoretical limit, HCI sensitivity scales with the fourth power of the telescope diameter. It makes the upcoming ELTs potentially hundreds of times more sensitive than current ground-based telescopes. This will be transformational - no longer will HCI be restricted only to young stellar systems found at tens to hundreds of parsecs, but will also be sensitive to cool, mature planets around nearby stars. Since angular resolution is inversely proportional to distance, our nearest neighbours will be the most valuable targets. An exciting preview is the JWST/MIRI detection[11] of the gas giant Epsilon Indi Ab, a cold (275 K), mature, 6 Jupiter-mass planet in a 15 au orbit. With hindsight it is visible in hours of archival VLT/VISIR data. METIS will see it with ease in a matter of minutes.

**The HCI landscape in 2035**

The first generation of instrumentation on the ELT will make use of the strong power-law scaling of its sensitivity for direct imaging. Each of the instruments will have some form of high-contrast imaging mode. However, most, if not all, of these instruments were not originally developed with high-contrast imaging in mind. While these instruments will outperform the current generation, they are far from ideal from a high-contrast imaging perspective. The planned Planetary Camera Spectrograph[12] (PCS), and Giant Magellan Adaptive Optics eXtreme[13] (GMagaO-X) instruments are specifically designed for high-contrast imaging for the ELT and GMT, respectively.

Some of the biggest challenges for the next-generation extreme adaptive optics systems are the high number of actuators and operational speed of adaptive mirrors, non-common path effects between wavefront and science channels, and time lag between wavefront measurement and correction. PCS and GmagAO-X for the GMT are expected to overcome these obstacles deploying multiple-stage AO systems, focal-plane wavefront sensing, and smart algorithms for predictive control –in combination with photon-noise limited post-processing techniques to achieve the fundamental sensitivity limit of the telescopes. High-resolution spectroscopy is an example to further filter out residual starlight (ANDES will be the first ELT instrument that utilizes this latter technique). This will result in the contrast limits of $10^{-7-9}$ required for temperate rocky-planet characterization around dozen(s) of nearby M-dwarfs. Going a step further, using coherence differential imaging[16] using all available sensors inside the instruments may allow us to go an order of magnitude deeper in contrast. This might make it possible for PCS and GMagAO-X to also detect the nearest rocky planets around sun-like stars. However, this limit will only be achieved if current theoretical ideas are actually turned into real designs for PCS and GMagAO-X. This is a great incentive to push strongly for hardware development, and software development, over a 5 and 10 year timeframe respectively.

**The future of ground-based exoplanet characterization beyond 2040**

Technologies for AO and coronagraphy are approaching the fundamental information limits set by physics, and PCS and GMAgAO-X are expected to operate close to this limit. Their final observing efficiency is then set by the photon flux of the objects that are observed, limited by the coherence time – the typical timescale of the turbulent atmosphere. AO needs to sense and correct the wavefront within this coherence time during which only a finite number of photons are available. In addition, post-processing methods need to predict the leaked stellar speckles to within the limit that is set by the photon flux. If these steps are achieved, the instruments cannot be improved any further. Laser-guide stars allow higher photon fluxes but must be naturally off-axis, probing a different column of atmospheric turbulence leading to worse sensitivity limits. This means that planet/star contrasts can only significantly improve by building an even larger telescope than the ELT or GMT.

Building an even bigger segmented/monolithic mirror is unlikely due to the ever increasing costs[17]. One way forward would be the development of a filled aperture coherent array of one to two meter dishes with small AO systems that are designed in a massively replicable way. These can then be coupled and combined with integrated photonic chips that currently already allow for hundreds of coupled modes. An approach like this would reduce costs and take advantage of economy of scales. A massive investment in coherent beam combining[18] is required over the next decade to make such a telescope possible.

**Detecting true Earth-twins** orbiting solar-type stars is one of the major goals of the exoplanet field. This requires a contrast level of $10^{-10}$ to $10^{-11}$. Even with near-optimal instrumentation, these will likely be out of realm of ground-based HCI systems. For this reason, the Habitable Worlds Observatory (HWO[14]) is positioned as NASA's future flagship space telescope specifically designed to identify and directly image >25 potential habitable worlds around sunlike stars. In addition, there are plans for the Large Interferometer For Exoplanets (LIFE) mission[15]. This concept consists of a nulling interferometer operating in the mid-infrared (4-18 μm), targeting exoplanet thermal emission. In particular a combination of HWO and LIFE will be very powerful for the detailed characterization of Earth-like planets.

We hope that a paradigm shift is arising that will be highly impactful for astronomers. Space industry is developing progressively larger launch vehicles with larger fairings for lower costs. These will provide an opportunity to launch larger telescopes for lower prices, possibly making single-science-case telescopes feasible with optimized telescope and instrument design that do not have to compromise on performance due to multiple conflicting science requirements.

One can argue whether in the age of HWO and/or LIFE, ground-based observatories could play a pioneering role in the characterisation of Earth-twins or not. It would require fundamentally new technologies to arise which are currently not foreseen. We may have to accept that at some point ground-based observatories will lose their competitiveness to space-based telescopes, as became the case for city-based observatories in the first half of the twentieth century.